\documentclass[twocolumn,showpacs,preprintnumbers,amsmath,amssymb,aps]{revtex4}

\usepackage{graphicx}
\usepackage{amssymb}
\usepackage{epstopdf}
\usepackage{bm}

\begin{document}
    
\title{Identifying the Unidentified Auger UHE Cosmic Rays with the Help of the Standard Model of Particle Physics}
 
\author{Frank~J.~Tipler}
\affiliation{Department of Mathematics and Department of Physics, Tulane University, New Orleans, LA 70118}

\date{\today}

\begin{abstract}
I have shown that if we assume that the Standard Model of particle physics and Feynman-Weinberg quantum gravity holds at all times, then in the very early universe, the Cosmic Background Radiation (CBR) cannot couple to right handed electrons and quarks.  If this property of CBR has persisted to the present day, the Ultra HIgh Energy Cosmic Rays (UHECR) can propagate a factor of ten further than they could if the CBR were an electromagnetic field, since most of the cross section for pion production when a UHECR hits a CBR photon is due to a quark spin flip, and such a flip cannot occur if the CBR photon cannot couple to right handed quarks.  The GZM effect will still reduce the number of UHECR, but UHECR can arrive from a distance of a redshift of up to $z=0.1$.  I show that taking this additional propagation distance into account allows us to identify the sources of  4 of the 6 UHECR which the Pierre Auger Collaboration could not identify, and also identify the source of the 320 EeV UHECR seen by the Fly's Eye instrument.   I suggest an experiment to test this hypothesis about the CBR.

\end{abstract}

\maketitle

\section{Introduction}
Astrophysicists should analyze experimental data assuming the validity of the extensively tested laws of physics.  The Standard Model (SM) of particle physics is one such law, but astrophysicists are not as familiar with the Standard Model as they are with the Maxwell equations and general relativity.  Thus astrophysicists typically do not take into account SM effects when analyzing data..  According to the SM, the electromagnetic field is not fundamental, but is instead composed of three true fundamental fields, an $SU(2)_L$ gauge field, a $U(1)$ gauge field, and a complex doublet Higgs scalar field.  I have shown \cite{Tipler2005} that quantum field theory does not permit  $U(1)$ radiation to exist  in the very early universe.  Intitially, in fact, the Cosmic Background Radiation (CBR) must have been entirely $SU(2)_L$ radiation. That is, initially, the CBR could only couple to left-handed fermions; this is the meaning of the subscript $L$.  If this property of the initial CBR has persisted to the present day, then the CBR would consist of pseudo-photons, just like photons, but unable to couple to fermions of right-handed chirality.  

This would mean that UHECR could propagate a factor of ten further through the CBR, since 90\% of the cross-section for pion production in the collision between a UHECR proton and a CBR photon is due to a quark spin flip, and a pseudo-photon cannot generate a quark spin flip.  The GZK effect would still exist --- the pion production cross-section is still non-zero --- and it would still make its appearance at the energy predicted by Greisen \cite{Greisen} and by Zatsepin and Kuz'min \cite{ZK}.  But higher energy cosmic rays would still be able to propagate through the CBR, from a distance as great as $z=0.1$.  I shall use this fact to show that all but two of the ``unidentified'' UHECR in the recently published data obtained by the Pierre Auger Observatory can also be associated with Active Galactic Nuclei (AGN).  I propose that within $3^\circ$ of the arrival direction of the two remaining unidentified UHECR, there will be found a previously unknown AGN.

\section{Summary of SM Cosmology}

Forty years of ever more precise observations have shown that the CBR has a Planckian distribution.  Most astrophysicists assume that a Planckian distribution means a field in thermal equilibrium, but this is false: a Planckian distribution need not arise from thermal process.  A Planckian distribution can also be a reflection of spacetime symmetries, as it is in the case of Hawking radiation in black hole spacetimes.  I have shown \cite{Tipler2005} that a quantized gauge field in a flat Friedmann universe necessarily has a Planckian distribution, with ``pseudo-temperature'' proportional to $1/R$, where $R$ is the scale factor of the Friedmann universe.  It is thus possible that the pure $SU(2)_L$ gauge field in the very early universe survived to the present (combined with the Higgs vacuum), in which case the CBR would not be electromagnetic thermal radiation, but instead would be radiation that  is missing the $U(1)$ piece, which would mean that the CBR would not couple to right-handed protons. The UHECR observations over the past few decades, including the Auger observations, indicate that this is the case.

 A central theorem of quantum field theory, the Bekenstein Bound, requires isotropy and homogeneity at the Planck time \cite{Bekenstein1989}.  The Bekenstein Bound also requires zero entropy and hence zero temperature at the Planck time.  Eddington \cite{Eddington1931} and Lema\^itre \cite{Lemaitre1931} in 1931, and Feynman \cite{Feynman} in 1963, argued that the only natural initial condition for the universe was one of zero entropy and zero temperature, so Bekenstein confirms their hypothesis.  Zero temperature forces the Sakarov conditions to be obeyed in the early universe, even with the Planck distribution: the Planck spectrum comes from the Friedmann geometry, and not thermal equilibrium.  Having a Planckian but non-thermal CBR in the very early universe allows the SM to explain why there is more matter than antimatter in the universe. 

The Bekenstein Bound picks out the $SU(2)_L$ field as the only SM field that can exist at the Planck time.  Furthermore, this $SU(2)_L$ field must be self-dual, so it will force tunneling between the vacua of the electroweak force, generating only particles, and not anti-particles.  A pure self dual $SU(2)_L$ at zero temperature field in the very early universe thus gives a new mechanism for Standard Model baryogenesis, and I have shown \cite{Tipler2005} that this  mechanism naturally generates the observed photon to baryon ratio.  Furthermore, the created baryons and leptons (baryon number minus lepton number, $B - L$, is conserved in SM baryogenesis) themselves break the homogeneity that the Bekenstein Bound requires at the Planck time, and in a spacetime that is very close to flat, generates the observed flat perturbation spectrum at the observed magnitude. 

Feynman and Weinberg showed many years ago that there is a unique renormalizable quantum theory of gravity for a spin two field, and we know gravity must be spin two.  I  have shown elsewhere \cite{Tipler2005} that with the appropriate cosmological boundary conditions \cite{Tipler2007}, the Feynman-Weinberg quantum theory of gravity is not only renormalizable, but term by term finite, and that the same mechanism that makes the theory term by term finite also forces the power series in the coupling constants to converge.  Applying this theory of quantum gravity to the pre-Planck time early universe, we find that the wave function of the universe must have been a delta function at the initial singularity.  An initial delta function wave function would {\it explode} outward, forcing the universe to be essentially flat, and the mechanism causing spatially flatness would thus be wave packet spreading, a well-known quantum process.  The universe's spatial flatness would thus be due to a kinematical quantum mechanical mechanism, and not a finely tuned dynamical mechanism.  A kinematical mechanism does not need fine tuning, but the inflation mechanism is known to require it.  Taking the SM into  account, we see that an inflationary era is not necessary to explain {\it any} cosmological observation.  

In particular, the SM, when combined with standard quantum mechanics, explains the observed isotropy, homogeneity, and spatial flatness of the universe, as well as the observed photon to baryon ratio, the observed excess of matter over anti-matter, and the Harrision-Zel'dovich perturbation spectrum.  And finally, it explains why we see the CZM  cut-off, and yet we also see UHECR with energies beyond the CZM cut-off.

\section{Analysis of Pierre Auger Data}

There were 27 events reported in the Pierre Auger Data \cite{Auger2}.  The Pierre Auger  Collaboration identified \cite{Auger1} a low redshift AGN or quasar within $3^\circ$ of the event direction for 21 of these.  By ``low redshift'' is meant $z\leq 0.01$, sufficiently close to the Earth to propagate through the CBR according to conventional GZK theory.  The Collaboration was cautious, but I completely agree with their obvious unstated opinion that these 21 events probably originated in the objects pointed out by the Collaboration.  However, 6 events were unidentified, and of  these unidentified events, 5 had energies above the median energy for the 27 reported event.  I find it highly significant that the unidentified events were clustered at the higher end of the observed energies.  The six unidentified events are listed in Table 1, together with the unidentified highest energy UHE cosmic ray ever observed, a 320 EeV event detected by Fly's Eye \cite{Sommers}.  In Table 1, the events are listed in increasing order of the energy.  

I used the online VizieR data base to rescan the locations of the unidentified events in the 2006 (12th edition) of the Veron + list of Quasars and Active Galactic Nuclei, looking for quasars and AGN with a redshift up to $z=0.1$, as allowed by the possibility that the CBR would not couple to right-handed protons, and hence could propagate to Earth from this distance.  I found acceptable sources within $3^\circ$ for 4 of the 6 unidentified sources.  For the other 2, I could find no quasar or AGN at any redshift within $3^\circ$ of the reported position of these event.  I list in Table 1 the two still unidentified objects with an asterisk, and the angular distance of the nearest quasar or AGN. In addition, I point out in Table 1 that there is also an acceptable possible source for the 320 EeV Fly's Eye event, a source discovered in 1994 by Elbert and Sommers, who rejected this source since its redshift is $0.020$, a factor of 2 too high to be the source by conventional GZK theory, but well within the limit allowed by the Standard Model theory I am proposing.  The SM theory also gives a source within $2^\circ$ for the 148 EeV event observed by Auger, a source too far away for conventional theory to explain, and in any case 148 EeV is too high an energy according to conventional theory.

\begin{table}
\caption{\label{tab:UHECR} List of Pierre Auger UHECR with Sources Unidentified.  The 320 EeV event is the Fly's Eye observation. Two events with still unidentified sources are mark with an asterisk.  For these two events are listed the closest AGN/quasars of {\it any} redshift.   Celestial coordinates are J2000.}
\begin{ruledtabular}
\begin{tabular}{lccccc}
E (EeV) &RA &Dec &{\rm suggested source} & {\rm redshift} & {\rm distance} \\
\hline
64*	 & $219.3^\circ$  	      &$-53.8^\circ$ &{\rm PKS 1427-448} &$z=?$ & $9^\circ$\\
70							& $267.1^\circ$  	      &$-11.4^\circ$ & {\rm 1H 1739-126} & $0.037$ &$3^\circ$\\
79 						& $201.1^\circ$      &$-55.3^\circ$& {\rm IRAS 13120-5453} &$0.031$ &$2^\circ$\\
80								&$185.4^\circ$      &$-27.9^\circ$&{\rm ESO 505-IG031} &$0.039$ &$2^\circ$\\
84*					&$114.6^\circ$      &$-43.1^\circ$&{\rm IRAS 07245-3548} &$0.029$ &$8^\circ$\\
148				&$192.7^\circ$     & $-21.0^\circ$&{\rm ESO 575-IG016} &$0.023$&$2^\circ$\\
320				&$85.2^\circ$     &$+48.0^\circ$ &{\rm MCG} $+08.11.011$  &$0.020$ &$3^\circ$ \\
\end{tabular}
\end{ruledtabular}
\end{table}

\section{Discussion}

Two events in Table 1 are marked with an asterisk.  These events are still unidentified, because there are no quasars or AGN at {\it any} redshift sufficiently close in angular position to be plausible sources.  Listed in the table are the closest quasars or AGN to the two unidentified objects.  I suggest that the region within $3^\circ$ of the positions of these two unidentified UHECR events be rescanned for quasars and AGN.  I predict that a new, previously undetected quasar or AGN will be found there.  If so, then we would then have a complete theory of  UHECR:  most come from nearby AGN, sufficiently close to not be stopped by the GZK effect.  But a few events will come from distances beyond the GZK cut-off.  The GZK cut-off is still there --- the threshold for pion production is unchanged --- but the cross-section reduction allowed by a CBR that cannot couple to right-handed proton allows UHECR from up to $z=0.1$.   (Another possibility is that the two unidentified events were UHECR with energies initially much higher, but which lost energy in a glancing collision with a CBR pseudo-photon, a collision that substantially changed the proton's propagation direction.  I know of no way to test this possibility without more UHECR data.)

When the GZK effect was first discovered,  one UHECR had been seen with an energy above 100 EeV, and Greisen himself wrote \cite{Greisen} in his paper on the GZK effect that it was surprising such a cosmic ray could exist at all.  Seeing these few events beyond the GZK cut-off is evidence that the CBR indeed cannot couple to right-handed protons.  Such a CBR would imply the GZK cut-off --- which has now been seen --- but would also allow the existence of truly ultra high energy cosmic rays --- which have also been seen.

If the CBR is an $SU(2)_L$ gauge field combined with the Higgs vacuum, and not a complete electromagnetic field, then it cannot couple to right-handed electrons either.  Thus we would expect CBR pseudo-photons to show substantially less Sunyaev-Zel-dovich effect that conventional theory would predict, as I pointed out in \cite{Tipler2005}.  This has now been seen \cite{Lieuetal06}.

Since these two completely different types of observations both indicate that the CBR may be composed  of pseudo-photons rather than photons, I suggest that this hypothesis be tested directly, as I have described in \cite{Tipler2005}.  However, I did not emphasize in \cite{Tipler2005} the importance of the ``active antenna effect,'' in reducing the size of the effect of the pseudo-photon nature of the CBR. Basically, it must be kept in mind that a pseudo-photon incident on a collection reflector will be absorbed and re-emitted --- this is the actual mechanism of all ``reflection'' --- but the re-emitted particle will be a photon, not a pseudo-photon.  This is a problem that can be circumvented.
\par
I  thank Paul Sommers and Alan A. Watson for some very helpful comments on the Pierre Auger experiment, and for directing me to the published data on the arXiv.  I am also grateful to Bruce Partridge for pointing out to me that the exponent of the last factor of equation (79) of \cite{Tipler2005} should be 1/4 rather than 1/2.








\begin{thebibliography}{99}


\bibitem{Auger1} Abraham, J. et al (Pierre Auger Collaboration) 2007a, Science 318, 938.

\bibitem{Auger2} Abraham, J. et al (Pierre Auger Collaboration) 2007b, arXiv:0712.2843v (17 December 2007).

\bibitem{Bekenstein1989} Bekenstein Jacob D 1989 ``Is the Cosmological Singularity Thermodynamically Possible?'' {\it Int. J. Theo. Phys.} {\bf 28} 967--981.

\bibitem{Sommers}Elbert, J. W., Sommers 1994 arXiv:astro-ph/9410069 v1 (20 October 1994).

\bibitem{Eddington1931}Eddington, Arthur S. 1931, ``The End of the World: From the Standpoint of Mathematical Physics,'' {\it Nature} {\bf 127} 447--453.

\bibitem{Feynman} Richard Feynman 1963 {\it The Feynman Lectures on Physics, Volume I} (New York: Addison-Wesley), Section 46-5.

\bibitem{Greisen}Greisen K. 1966 Phys. Rev. Lett. 16, 748.

\bibitem{Lemaitre1931}Lema\^itre, Georges, 1931 ``The Beginning of the World from the Point of View of Quantum Theory.''  {\it Nature} {\bf 127} (\#3210) 706.

\bibitem{Lieuetal06}Lieu, Richard, Jonathan P.D. Mittaz, and Shuang-Nan Zhang 2006 ``The Sunyaev-Zel'dovich Effect in a Sample of 31 Clusters --- a Comparison Between the X-ray Predicted and WMAP Observed CMB Temperature Decrement,'' {\it Astrophysical Journal} {\bf 648} 176--181.  I am grateful to Gary Hinshaw for this reference.

\bibitem{Tipler2} Tipler, F. J. 1994, {\it The Physics of Immortality}, Doubleday, New York

\bibitem{Tipler2005}Tipler, Frank J. 2005, ``The Structure of the World from Pure Numbers''  {\it Rep. Prog Phys.} {\bf 68}  897--964.

\bibitem{Tipler2007}Tiper, Frank J., Jessica Graber, Matthew McGinley, Joshua Nichols-Barrer, and Christopher Staecker, 2007, ``Closed Universes with Black Holes but No Event Horizons as a Solution to the Black Hole Information Problem''  {\it Mon. Not. Royal Astron. Soc.} {\bf 379}, 629--640.

\bibitem{ZK}Zatsepin, G.T., Kuz'min 1966 Sov. Phys. JETP Lett. 4, 78.




\end{thebibliography}
\end{document}